\documentclass[12pt]{article}
\usepackage{a4wide}

\usepackage[dvipdfmx,hiresbb]{graphicx} 
\usepackage{mediabb}                     

\usepackage{color}
\newcommand{\magenta}[1]{#1}
\newcommand{\blue}[1]{#1}
\definecolor{refkey}{rgb}{1,0,1}
\definecolor{labelkey}{rgb}{1,0,1}

\newcommand{\al}[1]{\begin{align}#1\end{align}}

\newcommand{\ab}[1]{\left|#1\right|}
\newcommand{\paren}[1]{\left(#1\right)}
\newcommand{\fn}[1]{\!\left(#1\right)}

\usepackage{amsmath,amssymb}
\newcommand{\df}{\text{d}}

\usepackage{epsf}

\newcommand{\cred}[1]{{\color{black}#1}}

\newcommand{\GeV}{\ensuremath{\,\text{GeV} }}

\def\beq{\begin{eqnarray}}
\def\eeq{\end{eqnarray}}

\newcommand{\bea}{\begin{eqnarray}}   
\newcommand{\eea}{\end{eqnarray}}
\newcommand{\bear}{\begin{array}}  
\newcommand {\eear}{\end{array}}
\newcommand{\bef}{\begin{figure}}  
\newcommand {\eef}{\end{figure}}
\newcommand{\bec}{\begin{center}}  
\newcommand {\eec}{\end{center}}

\begin{document}

\baselineskip 0.7cm

\begin{titlepage}

\begin{flushright}
KUNS-2511\\
OU-HET/825-2014\\
TU-981\\
IPMU 14-0292\\
\end{flushright}

\vskip 1.35cm
\begin{center}
{\Large \bf 
Topological Higgs inflation:\\
\large The origin of the  Standard Model criticality
}
\vskip 1.2cm
Yuta Hamada,$^{a}$
Kin-ya Oda$^{b}$
and
Fuminobu Takahashi$^{c,d}$

\vskip 0.4cm

{ \it $^a$ Department of Physics, Kyoto University Kyoto 606-8502, Japan}\\
{\it $^b$ Department of Physics, Osaka University, Osaka 560-0043, Japan}\\
{\it $^c$ Department of Physics, Tohoku University, Sendai 980-8578, Japan}\\
{\it $^d$ Kavli Institute for the Physics and Mathematics of the Universe (WPI), 
 TODIAS, University of Tokyo, Kashiwa 277-8583, Japan    }

\vskip 1.5cm

\abstract{
The measured values of the Higgs and top masses and of the strong gauge coupling constant point to the near-criticality of the Standard Model, where two vacua at the electroweak and Planck scales are 
quasi-degenerate.
We argue that the criticality is required by the occurrence of an eternal topological inflation 
induced by the Higgs potential. The role of this inflation is to continuously create sufficiently flat and homogeneous Universe, providing the necessary initial condition for the subsequent slow-roll inflation that generates the density perturbations of the right 
magnitude. 
While the condition for the topological Higgs inflation is only marginally satisfied in the Standard Model, 
it can be readily satisfied if one introduces the right-handed neutrinos and/or the non-minimal coupling to gravity;
currently unknown quantum gravity corrections to the potential may also help.
We also discuss the $B-L$ Higgs inflation as a possible origin of the observed density perturbations.
Its necessary initial condition, the restored $B-L$ symmetry, can be naturally realized by the preceding topological Higgs inflation.
}
\end{center}
\end{titlepage}

\setcounter{page}{2}

\section{Introduction}
The discovery of the Standard Model (SM) Higgs boson only so far at the 
LHC~\cite{Aad:2012tfa, Chatrchyan:2012ufa} already provided us
with various implications for theory beyond the SM as well as the Universe we live in. 
In particular,  the experiments seem to suggest a special structure of the vacuum.

The measured Higgs boson mass about $125$\,GeV~\cite{ATLAS2014,CMS2014}
implies that the SM could be valid all the way up to the Planck scale. 
If so, the Higgs potential \magenta{may have} another minimum around the Planck scale, \magenta{depending on the top quark mass};
\magenta{see e.g.\ Refs.~\cite{Holthausen:2011aa,Bezrukov:2012sa,Degrassi:2012ry,Alekhin:2012py,Masina:2012tz,Hamada:2012bp,Jegerlehner:2013cta,Buttazzo:2013uya,Branchina:2013jra,Kobakhidze:2014xda,Spencer-Smith:2014woa,Branchina:2014usa,Branchina:2014rva} for the latest analyses}.
If the extra minimum is the global one, our electroweak
vacuum is unstable and  decays through quantum tunneling processes with a finite lifetime.
\magenta{In contrast, nature may realize} a critical situation where the two minima are degenerate in energy.
Froggatt and Nielsen focused on this special case, the so-called  Higgs criticality; they  provided a theoretical argument to support this case, \magenta{the multiple point criticality principle~\cite{Froggatt:1995rt,Nielsen:2012pu}}.\footnote{
\magenta{See e.g.\ Section 5.A.1 of Ref.~\cite{Hamada:2014wna} for a list of other possibilities to realize the (near) criticality.}
}
The near-critical behavior of the Higgs potential is a puzzle and has led to much excitement in \magenta{the} context of Higgs inflation~\cite{Bezrukov:2007ep,Masina:2011aa,Masina:2011un,Allison:2013uaa,Hamada:2013mya,Cook:2014dga,Hamada:2014iga,Bezrukov:2014bra}.

In this Letter we point out that the near-criticality can be understood if the Universe experiences
\blue{eternal} topological inflation~\cite{Linde:1994hy, Linde:1994wt,Vilenkin:1994pv} induced by the Higgs field
at a very early stage.
We study the condition for the topological Higgs inflation to occur in the SM and what kind of extensions
can relax the condition. As we shall see shortly, \blue{while} the condition is only marginally 
satisfied in the SM, it can be readily satisfied if one extends the SM
by introducing heavy right-handed neutrinos and/or a non-minimal coupling to gravity.

The topological Higgs inflation may be thought of as one of the variants of the Higgs inflation,
but it is different in the following aspects. First, the topological inflation is free of the initial
condition problem. If the Universe begins in a chaotic state at an energy close to the Planck scale,
the Higgs field may take various field values randomly up to the Planck scale or higher~\cite{Linde:1983gd}. 
As the Universe expands, the energy density decreases and the Higgs field finds itself
either larger or smaller than the critical field value corresponding to the local maximum, and gets
trapped in one of the two degenerate vacua with a more or less equal probability. 
This leads to  formation of domain walls separating the two vacua. Interestingly,
then, eternal inflation could take place inside the domain walls, if the thickness of the domain walls is
greater than the Hubble radius~\cite{Linde:1994hy,Vilenkin:1994pv}. 
In this sense, no special fine-tuning of the
initial position of the inflaton is necessary for the inflation to take place. Specifically, 
the topological inflation  occurs if the two minima are separated by more than the Planck 
scale,  which was also confirmed by numerical calculations~\cite{Sakai:1995nh}. 
Secondly, the magnitude of density perturbations generated
by topological Higgs inflation tends to be too large to explain the observed CMB temperature fluctuations. 
We need therefore another inflation after the end of the topological Higgs inflation, and we will
return to this issue later \blue{in this Letter}. 
\blue{Thus, the role of the topological Higgs inflation is to continuously 
create sufficiently flat Universe, solving the so-called longevity problem of inflation with a Hubble parameter much 
smaller than the Planck scale~\cite{Linde:1983gd,Izawa:1997df}; the Universe must be sufficiently flat and
therefore long-lived so that the subsequent slow-roll inflation with a much smaller Hubble parameter can take place.}

\section{Topological Higgs inflation in SM}


For a Higgs field value much larger than the electroweak scale, the effective potential is \blue{approximately} 
given by\footnote{
Here we can safely treat the Higgs inflation as a single-field model. See footnote 5 of Ref.~\cite{Hamada:2014iga} and Refs.~\cite{Greenwood:2012aj,Kaiser:2013sna}. Eq.~\eqref{effective potential} can be obtained from the one-loop effective potential, including the loops from the NG bosons.
}
\al{
V={1\over 4}\lambda_\text{eff}\fn{\varphi}\varphi^4,
	\label{effective potential}
}
where the effective coupling is expanded around its minimum as
\al{
\lambda_\text{eff}\simeq\lambda_\text{min}+{\beta_2\over\paren{16\pi^2}^2}
\left(\ln{{\varphi\over\mu_\text{min}}}\right)^2,
	\label{lambda_eff}
}
with $\beta_2\simeq0.5$; see e.g.\ Refs.~\cite{Hamada:2014iga,Hamada:2014wna}.

\begin{figure}[tn]
\begin{center}
\hfill
\includegraphics[width=.6\textwidth]{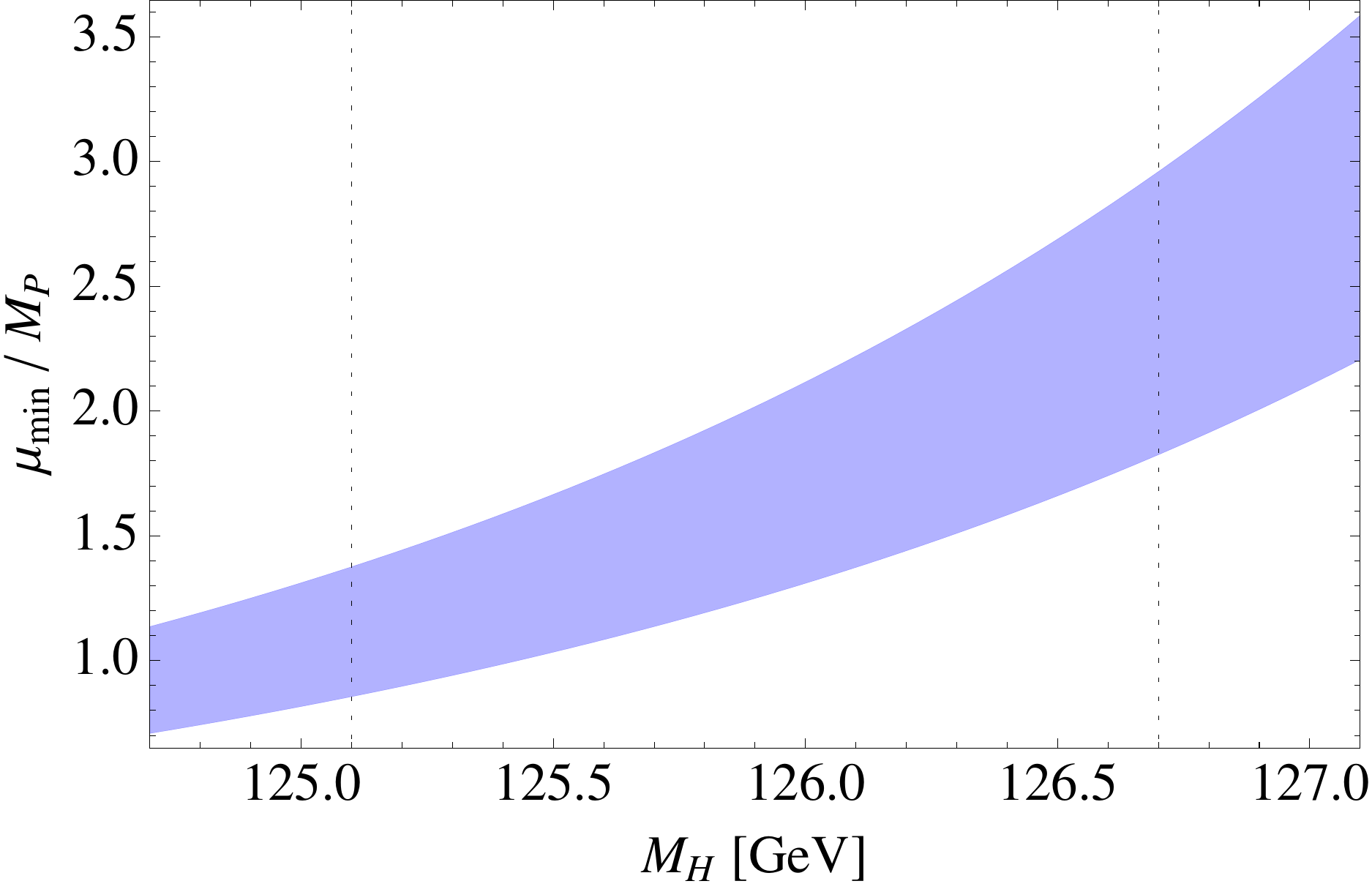}
\hfill\mbox{}\\
\caption{The scale $\mu_\text{min}$ that gives the minimum of the effective coupling $\lambda_\text{eff}$, under the criticality condition $\lambda_\text{min}:=\lambda_\text{eff}\fn{\mu_\text{min}}=0$, as a function of $M_H$. The band width corresponds to the 2$\sigma$ deviation of $\alpha_s$, with 1$\sigma$ being given by $\alpha_s=0.1185\pm0.0006$~\cite{PDG2014}.
The dotted lines shows the 2$\sigma$ band for $M_H$, with 1$\sigma$ being given by $M_H=125.9\pm0.4\GeV$~\cite{PDG2014}.
}
\label{SMvalue}
\end{center}
\end{figure}

Following the multi-point criticality principle, we assume $\lambda_\text{min}=0$.
The derivatives of the potential read
\al{
V_\varphi
	&=
{\beta_2\over\paren{16\pi^2}^2}\ln{{\varphi\over\mu_\text{min}}}
\left(
\ln{{\varphi\over\mu_\text{min}}}+{1\over2}
\right)
\varphi^3,\\
V_{\varphi\varphi}
	&=
{\beta_2\over2\paren{16\pi^2}^2}
\paren{1+\ln{\varphi\over\mu_\text{min}}}
\paren{1+6\ln{\varphi\over\mu_\text{min}}}
\varphi^2.
}
Note that $\varphi=\mu_\text{min}$ also gives the minimum of the potential when the criticality condition $\lambda_\text{min}=0$ is met.
The local maximum of the potential is at
\al{
\varphi_t=e^{-1/2}\mu_\text{min}.
}
\blue{Let us take a domain wall separating the two degenerate minima at $\varphi = \mu_\text{min}$ and $\varphi \approx 0$.}\footnote{
\blue{ Here and in what follows we neglect the electroweak scale vacuum expectation value compared to the Planck scale.}}
The typical thickness of the domain wall is given by~\cite{Sakai:1995nh}
\al{
\delta
	\simeq	\ab{V_{\varphi\varphi}\fn{\varphi_t}}^{-1/2}
	=	{16\pi^2\over\mu_\text{min}}\sqrt{2e\over\beta_2}.
}
\blue{If the thickness is greater than the Hubble radius, the domain wall will expand, and
topological inflation takes place~\cite{Linde:1994hy,Vilenkin:1994pv}.}
The Hubble parameter at the maximum \blue{(i.e. around the center of the domain wall)} is
\al{
H^2
	=	{V(\varphi_t)\over3M_P^2}
	=	{\beta_2\over\paren{16\pi^2}^2}{\mu_\text{min}^4\over48e^2M_P^2}
	=	\paren{2.4\times 10^{-4}M_P}^2\paren{{\beta_2\over0.5}}\paren{{\mu_\text{min}\over M_P}}^4,
}
\magenta{where $M_P\simeq2.4\times10^{18}\GeV$ is the (reduced) Planck scale.}
The condition for the domain wall to expand is~\cite{Sakai:1995nh}
\al{
H \delta
	=	\sqrt{V(\varphi_t)\over3M_P^2\ab{V_{\varphi\varphi}(\varphi_t)}}
	=	{1\over\sqrt{3\ab{\eta(\varphi_t)}}}
	\gtrsim
		0.48,	\label{H delta}
}
namely,
\al{
\ab{\eta(\varphi_t)}
	&\lesssim	1.4,
		\label{eta condition}
}
\blue{where $\eta \equiv M_P^2 V_{\varphi \varphi}/V$ is one of the slow-roll parameters.}
For the SM, this condition reads
\al{
\mu_\text{min} \gtrsim 3.9 M_P.
	\label{SM condition}
}
\blue{Note that the prefactor of this condition may contain an ${\cal O}(1)$ uncertainty, because it is derived for a $Z_2$ symmetric potential,
and the precise condition depends on the detailed shape of the inflaton potential.}
In Fig.~\ref{SMvalue}, we plot the scale $\mu_\text{min}$ at which the effective coupling $\lambda_\text{eff}$ takes its minimum value, under the criticality condition $\lambda_\text{min}:=\lambda_\text{eff}(\mu_\text{min})=0$; \magenta{in the computation, we have used the two-loop renormalization group equations and the one-loop effective potential in the Landau gauge}. We see that $\mu_\text{min}$ is indeed \magenta{of the order of} the Planck scale, \magenta{given} \blue{the measured mass of the Higgs boson. The condition (\ref{SM condition}) may be satisfied taking account of the uncertainty.}

If one wanted to use this potential to generate the observed density fluctuation, the potential height must satisfy~\cite{Ade:2013uln}
\al{
V(\varphi_t)
	&=	{\beta_2\over\paren{16\pi^2}^2}{\mu_\text{min}^4\over16e^2}
	<	\paren{1.94\times 10^{16}\GeV}^4{r_*\over0.12},
}
namely,
\al{
\mu_\text{min}<0.4M_P\paren{r_*\over0.12}^{1/4}\paren{0.5\over\beta_2}^{1/4},
}
\blue{which is inconsistent with the condition~\eqref{SM condition}.}
Therefore, one cannot use this topological inflation directly to generate the observed density fluctuation, \blue{
and we need another inflation after the topological Higgs inflation.}
The minimum scale $\mu_\text{min}$ can be \blue{larger} if one includes the right-handed neutrino~\cite{Haba:2014zda,Hamada:2014xka}.
If one embeds the SM in string theory, the string states may also change $\mu_\text{min}$; \magenta{see also Ref.~\cite{Hebecker:2013lha} that considers the near-criticality in string theory context}.

\section{Non-minimal coupling to gravity}
We may add the non-minimal coupling $\xi |H|^2 R$.\footnote{\blue{
The topological inflation  in the Starobinsky model was considered in Ref.~\cite{Kamada:2014gma}.}
} The Einstein frame potential then becomes 
\al{
U={\lambda_\text{eff}\fn{\mu}\over4}{\varphi^4\over\paren{1+\xi \varphi^2/M_P^2}^2}.
}
There are two prescriptions for the choice of the scale $\mu$:
\al{
\mu&={\varphi\over\sqrt{1+\xi \varphi^2/M_P^2}}
	&	&\text{(prescription I)},\\
\mu&=\varphi
	&	&\text{(prescription II)}.
}
In the Einstein frame, the canonically normalized field \blue{ $\chi$ is related to $\varphi$ as}
\al{
{\df\chi\over \df\varphi}
	&=	{\sqrt{1+\xi{\varphi^2\over M_P^2}+6\xi^2{\varphi^2\over M_P^2}}\over1+\xi{\varphi^2\over M_P^2}}.
}
\blue{In the following, let} us check whether the condition~\eqref{eta condition} can be satisfied in each prescription by
\blue{estimating the slow-roll parameter defined by}
\al{
\eta=M_P^2{U_{\chi\chi}\over U}.
	\label{eta in Einstein frame}
}

\begin{figure}[tn]
\begin{center}
\hfill
\includegraphics[width=.4\textwidth]{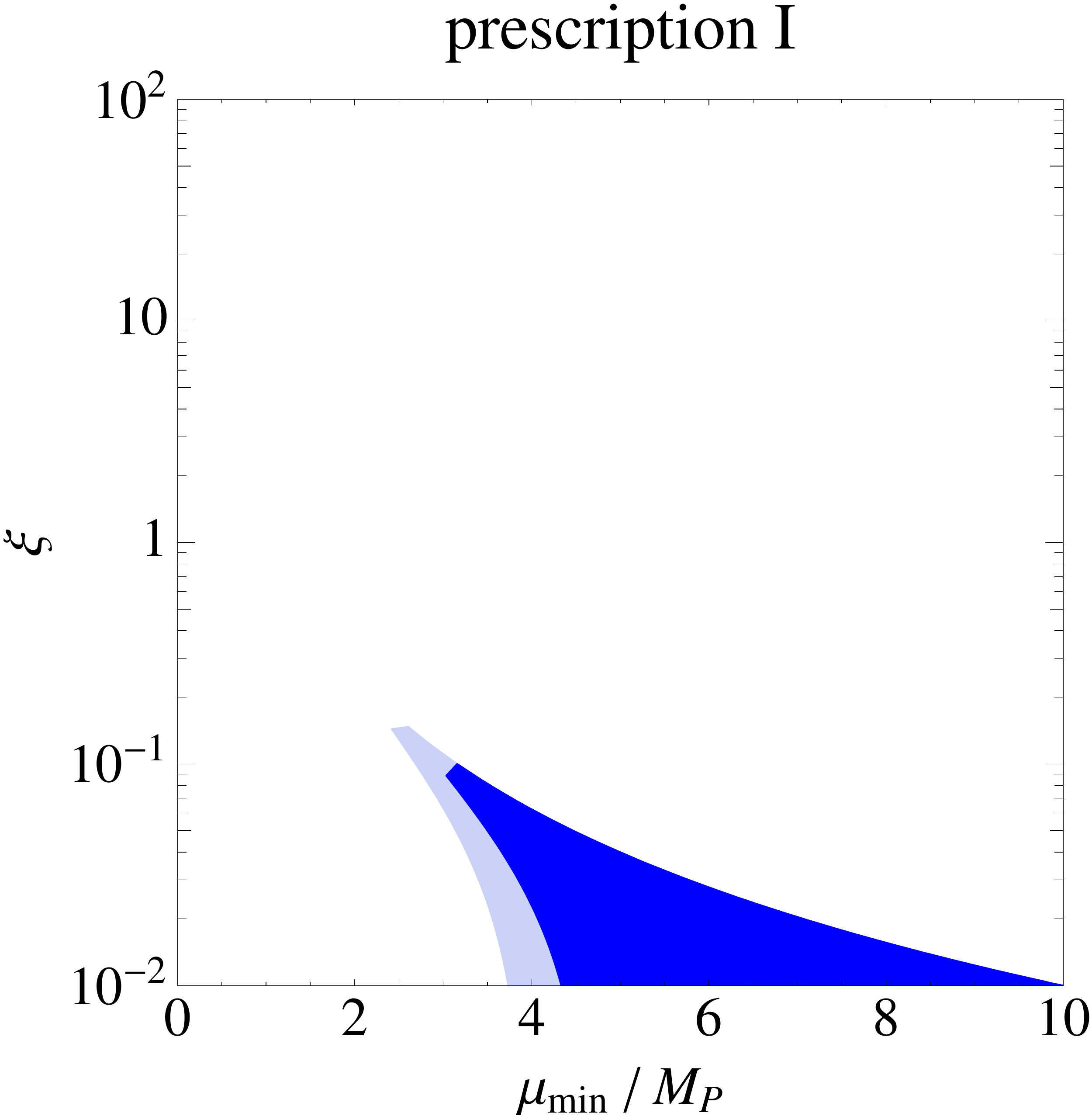}
\hfill
\includegraphics[width=.4\textwidth]{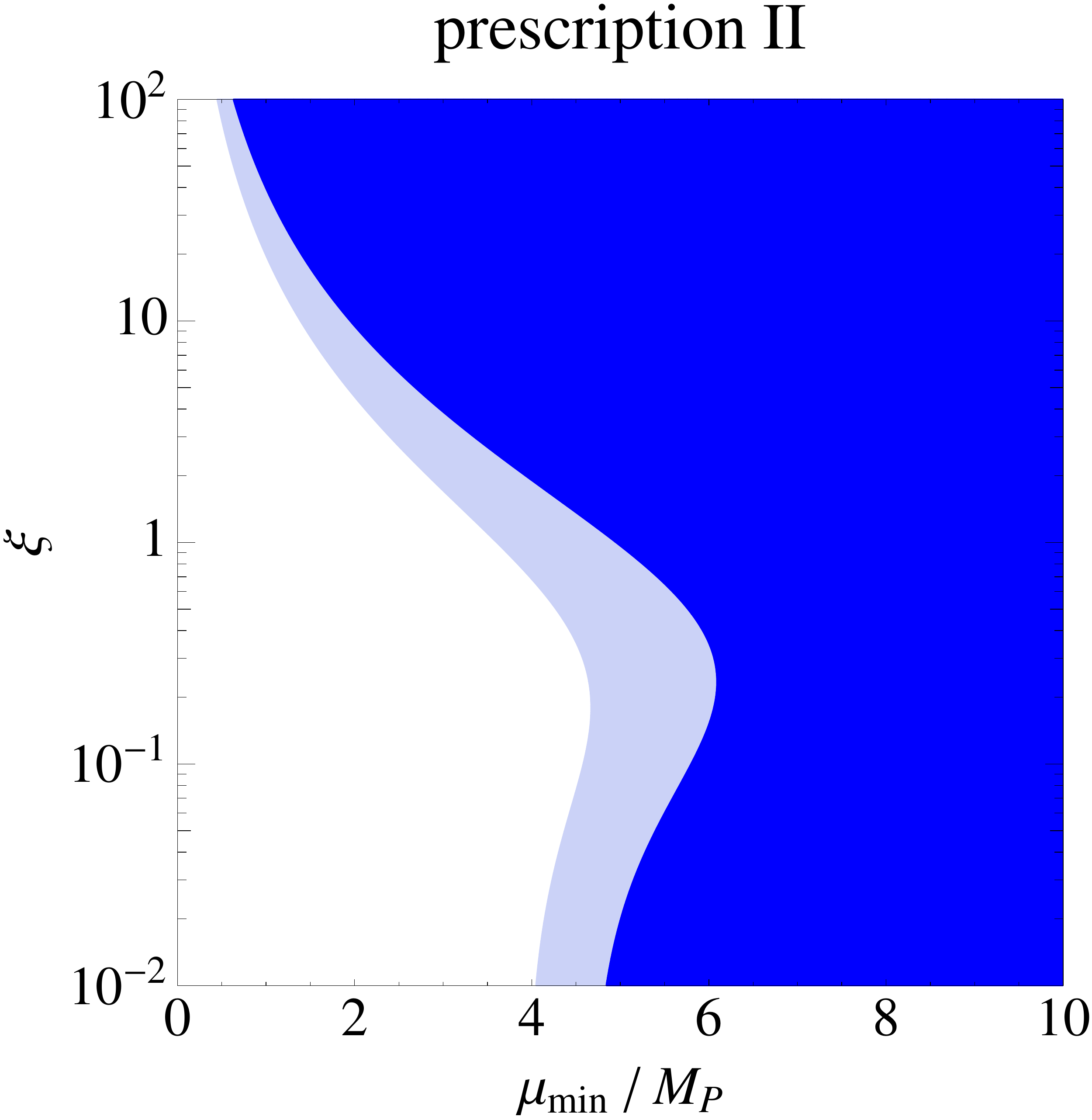}
\hfill\mbox{}\\
\caption{Allowed region in the prescription I (left) and II (right) in $\xi$ vs $\mu_\text{min}$ plane,
where dark (light) region satisfies $\ab{\eta(\varphi_\text{I})}<1$ (1.4) and $\ab{\eta(\varphi_\text{II})}<1$ (1.4), respectively. In the prescription I (left), $\mu_\text{min}<M_P/\sqrt{\xi}$ is also imposed.
}
\label{prescriptionI}
\end{center}
\end{figure}

\subsection{Prescription I}
The position of the local maximum and minimum of $U$, under the criticality condition $\lambda_\text{min}=0$, are respectively
\al{
\varphi_\text{I}
	&=	\mu_\text{min}{1\over\sqrt{e-\xi{\mu_\text{min}^2\over M_P^2}}}, &
\varphi_\text{min}
	&=	\mu_\text{min}{1\over\sqrt{1-\xi{\mu_\text{min}^2\over M_P^2}}},
}
\blue{where $\mu_\text{min}$ is defined by Eq.~\eqref{lambda_eff} even in  presence of $\xi$.}
We see that $\mu_\text{min}<M_P/\sqrt{\xi}$ is required in order to allow the local minimum $\varphi_\text{min}$.
At $\varphi_\text{I}$, the slow-roll parameter becomes
\al{
\eta
	&=	-8{M_P^2\over \mu_\text{min}^2}{\paren{e-\xi{\mu_\text{min}^2\over M_P^2}}^2 \over e+6\xi^2{\mu_\text{min}^2\over M_P^2}}.
		\label{eta in I}
}
In the left panel of Fig.~\ref{prescriptionI}, we plot the allowed region \blue{where \magenta{the condition~\eqref{eta condition} with Eq.~\eqref{eta in I}} is satisfied.
We can see from the figure that there is an allowed region with $\xi \sim 0.1$ and $\mu_\text{min} \sim 3 M_P$, \magenta{which lies in the} reasonable range of $\mu_\text{min} \simeq (1\text{--}3)M_P$ in the SM, shown in Fig.~\ref{SMvalue}.
}

\subsection{Prescription II}

In the prescription II, the local maximum of $U$, under the criticality condition $\lambda_\text{min}=0$, is located at
\al{
\varphi_\text{II}
	&=	{M_P\over\sqrt{\xi}}\sqrt{W\fn{\xi\mu_\text{min}^2\over e M_P^2}},
}
where $W$ is the Lambert function defined by $z=W(z)e^{W(z)}$.
The slow-roll parameter at this maximum is
\al{
\eta(\varphi_\text{II})
	&=	-{8M_P^2\over\varphi_\text{II}^2}{1+\xi{\varphi_\text{II}^2\over M_P^2}\over1+\xi\paren{1+6\xi}{\varphi_\text{II}^2\over M_P^2}}.
}
In the right panel of Fig.~\ref{prescriptionI}, we plot the allowed region.
\blue{We can see that there is a large allowed region for $\xi \gtrsim {\cal O}(1\text{--}10)$ 
for $\mu_\text{min} \simeq (1\text{--}3 )M_P$. We have checked that the typical value of $|\eta|$ is
 greater than $0.1$ for $\xi \lesssim 10^5$,
and therefore, it is difficult to account for the observed density perturbations \magenta{by this inflation alone}.}

\section{Discussion}

So far we have focused on the critical case where the two minima are degenerate. Successful topological 
Higgs inflation will be still possible even if the potential energy of the Planck-scale minimum is slightly higher or lower
than the electroweak minimum. If the degeneracy is largely broken, however, the trapping probability of the Higgs field
 will be significantly biased to the minimum with a lower energy. 
 This would reduce the number of domain-walls, especially that with an infinite length.
Also the domain walls become unstable and some of them will collapse before the topological inflation takes place.
Thus, we expect that the topological Higgs inflation will become less likely as the Higgs potential goes away from
the criticality.  We argue, therefore,  that the topological Higgs inflation which occurred at the very beginning of the
Universe could be the reason why the measured Higgs boson mass points to the near-criticality of the SM vacuum.
In order to estimate quantitatively to what extent the degeneracy can be broken, 
one must resort to numerical simulation, which is left for future work.

 \blue{As we have mentioned earlier,  while the topological Higgs inflation cannot account for the
 observed CMB temperature anisotropy, it creates sufficiently flat  and homogeneous Universe,
 and sets the required initial condition for the subsequent slow-roll inflation with a much smaller Hubble
 parameter~\cite{Linde:1983gd,Izawa:1997df}.}\footnote{\blue{It is interesting to study various extensions of the SM to see if
the topological Higgs inflation can generate the density perturbations of the right magnitude.}
} There is a variety of possibilities, but one of the 
interesting candidate is the inflation based on the $B-L$ Higgs inflation~\cite{Nakayama:2011ri,Nakayama:2011ys,Nakayama:2012dw,Barenboim:2013wra,Iso:2014gka}, where the Coleman-Weinberg potential gives a
dominant contribution to the tilt of the inflaton potential.\footnote{\cred{
This extension possibly changes the running of Higgs self coupling $\lambda$ due to the quartic coupling between the $B-L$ and SM Higgses. However, such a coupling does not affect the running of $\lambda$ if it is smaller than of order $10^{-2}$. Unless this coupling is extremely small, it helps to keep the $B-L$ Higgs field at its origin during 
the first inflation, setting the initial condition for the subsequent $B-L$ Higgs inflation.  Detailed study of this issue will be presented elsewhere.}}
The well-known problem of the density perturbations
in the Coleman-Weinberg inflation model can be avoided
if the $U(1)_{B-L}$ gauge coupling is small~\cite{Hawking:1982cz,Starobinsky:1982ee,Guth:1982ec}
 or if there is a cancellation between the gauge and neutrino Yukawa contributions~\cite{Ellis:1982ed,Nakayama:2011ri}.
There are a couple of interesting implications.  First, the inclusion of the right-handed neutrinos makes it easier to satisfy the condition for the topological Higgs inflation, \blue{since the $\mu_\text{min}$ becomes larger} \magenta{for the fixed $\lambda_\text{min}=0$.}\footnote{\cred{
The neutrino Dirac Yukawa couplings do not affect the Higgs potential significantly if they are  smaller than of order $0.1$; see Ref.~\cite{Hamada:2012bp}.
}}
\blue{With three right-handed neutrinos,
\magenta{adding the $U(1)_{B-L}$} is one of the plausible extensions of the SM as it is anomaly free.}
Secondly, the Higgs field will induce preheating as it oscillates about the origin after the 
topological inflation~\cite{Bezrukov:2008ut,GarciaBellido:2008ab}.
In order to estimate the reheating temperature precisely, one needs a detailed analysis of dissipation 
processes~\cite{Mukaida:2012bz}, but the reheating temperature is expected to be rather high. 
If so, the $U(1)_{B-L}$ symmetry is likely restored, setting the initial condition for the $B-L$ Higgs 
inflation.\footnote{During the topological Higgs inflation, the $B-L$ Higgs field may be at the origin or may develop a 
non-zero value. This affects the contribution of the neutrino Yukawa coupling to the SM Higgs potential,
 whose quantitative estimate will be left for future work.
}   \blue{Thus, the topological Higgs inflation provides the required initial condition for the $B-L$ Higgs inflation.}
Thirdly, the $B-L$ Higgs field will decay into right-handed neutrinos, whose CP-violating decay can generate
a right amount of the baryon asymmetry via leptogenesis~\cite{Fukugita:1986hr}. 

\section*{Acknowledgment}
\magenta{We are grateful to the Yukawa Institute for Theoretical Physics for the hospitality during the workshop PPP2014 (YITP-W-14-05), where the present work started, and to its participants for fruitful discussions.
We thank Hikaru Kawai for useful comments.} This work was \magenta{in part} supported by \magenta{the Grant-in-Aid from the JSPS Fellows No.\ 25.1107 [YH], the Grant-in-Aid for Scientific Research Nos.\ 23104009, 20244028, and 23740192 [KO]}, the Grant-in-Aid for Scientific Research on Innovative Areas (No.23104008 [FT]),  JSPS Grant-in-Aid for Young Scientists (B) (No.24740135 [FT]), Scientific Research (A) (No.26247042 [FT]), Scientific Research (B) (No.26287039 [FT]), \blue{and} Inoue Foundation for Science [FT].  This work was also supported by World Premier International Center Initiative (WPI Program), MEXT, Japan [FT].

\bibliographystyle{TitleAndArxiv}
\bibliography{Higgs}

\end{document}